\documentstyle[sprocl,epsfig]{article}

\def\be{\begin{equation}}
\def\ee{\end{equation}}
\def\bea{\begin{eqnarray}}
\def\eea{\end{eqnarray}}

\begin{document}

\title{PHENOMENOLOGY OF NEUTRINO OSCILLATIONS AT A NEUTRINO FACTORY}

\author{OSAMU YASUDA}

\address{Department of Physics,
Tokyo Metropolitan University \\
Minami-Osawa, Hachioji, Tokyo 192-0397, Japan
\\E-mail: yasuda@phys.metro-u.ac.jp}


\maketitle\abstracts{ It is shown in the three flavor framework
that neutrino factories enable us
to measure some of the oscillation parameters, such as
the sign of $\Delta m_{32}^2$, $\theta_{13}$, $\delta$.
Some efforts are made to determine the parameters (the muon energy
and the neutrino path length) of a neutrino factory
to optimize the signals.}

\section{Introduction}

There have been several experiments
\cite{homestake,Kamsol,SKsol,sage,gallex,Kamatm,IMB,SKatm,SKup,soudan2,lsnd}
which suggest neutrino oscillations.
It has been known in the two flavor framework that the solar neutrino deficit
can be explained by neutrino oscillation with the set of parameters
$(\Delta m^2_\odot,\sin^22\theta_\odot)\simeq$ $({\cal
O}(10^{-5}{\rm eV}^2),{\cal O}(10^{-2}))$ (small angle MSW solution),
$({\cal O}(10^{-5}{\rm eV}^2),{\cal O}(1))$ (large angle MSW
solution), or $({\cal O}(10^{-10}{\rm eV}^2),{\cal O}(1))$ (vacuum
oscillation solution), and the atmospheric neutrino anomaly
can be accounted for by $(\Delta m_{\mbox{\rm
atm}}^2,~\sin^22\theta_{\mbox{\rm atm}})\simeq (10^{-2.5}{\rm
eV}^2,1.0)$.
In the three flavor framework there are two independent mass squared
differences and it is usually assumed that these two mass differences
correspond to $\Delta m^2_\odot$ and $\Delta m_{\mbox{\rm atm}}^2$.
Throughout this talk I will assume three neutrino species which
can account for only the solar neutrino deficit and the atmospheric
neutrino anomaly \footnote{To explain the LSND
anomaly \cite{lsnd} one needs at least four neutrino species.}.
Without loss of generality I assume
$|\Delta m_{21}^2|<|\Delta m_{32}^2|<|\Delta m_{31}^2|$
where $\Delta m^2_{ij}\equiv m^2_i-m^2_j$.
The flavor eigenstates are related to the mass eigenstates by
$U_{\alpha j}$ ($\alpha=e,\mu,\tau$),
where $U_{\alpha j}$ are the elements
of the MNS mixing matrix U \cite{mns}:
\begin{eqnarray}
&{\ }&\left( \begin{array}{c} \nu_e  \\ \nu_{\mu} \\ 
\nu_{\tau} \end{array} \right)
=U\left( \begin{array}{c} \nu_1  \\ \nu_2 \\ 
\nu_3 \end{array} \right),\nonumber\\
U&\equiv&\left(
\begin{array}{ccc}
U_{e1} & U_{e2} &  U_{e3}\\
U_{\mu 1} & U_{\mu 2} & U_{\mu 3} \\
U_{\tau 1} & U_{\tau 2} & U_{\tau 3}
\end{array}\right)
=\left(
\begin{array}{lll}
c_{12}c_{13} & s_{12}c_{13} &  s_{13}e^{-i\delta}\nonumber\\
-s_{12}c_{23}-c_{12}s_{23}s_{13}e^{i\delta} & 
c_{12}c_{23}-s_{12}s_{23}s_{13}e^{i\delta} & s_{23}c_{13}\nonumber\\
s_{12}s_{23}-c_{12}c_{23}s_{13}e^{i\delta} & 
-c_{12}s_{23}-s_{12}c_{23}s_{13}e^{i\delta} & c_{23}c_{13}\nonumber\\
\end{array}\right).
\label{eqn:mns}
\end{eqnarray}
With the mass hierarchy $|\Delta m_{21}^2|\ll|\Delta m_{32}^2|$
there are two possible
mass patterns which are depicted in Fig. 1a and 1b,
depending on whether $\Delta m_{32}^2$ is positive or negative.
If the matter effect is relevant then  the sign of
$\Delta m_{32}^2$ can be determined by distinguishing neutrinos and
anti-neutrinos.
The Superkamiokande experiment uses water Cherenkov detectors
and events of neutrinos and anti-neutrinos are unfortunately
indistinguishable, so the sign of $\Delta m_{32}^2$ is unknown
to date.

\vglue -1.5cm 
\hglue 1.5cm
\epsfig{file=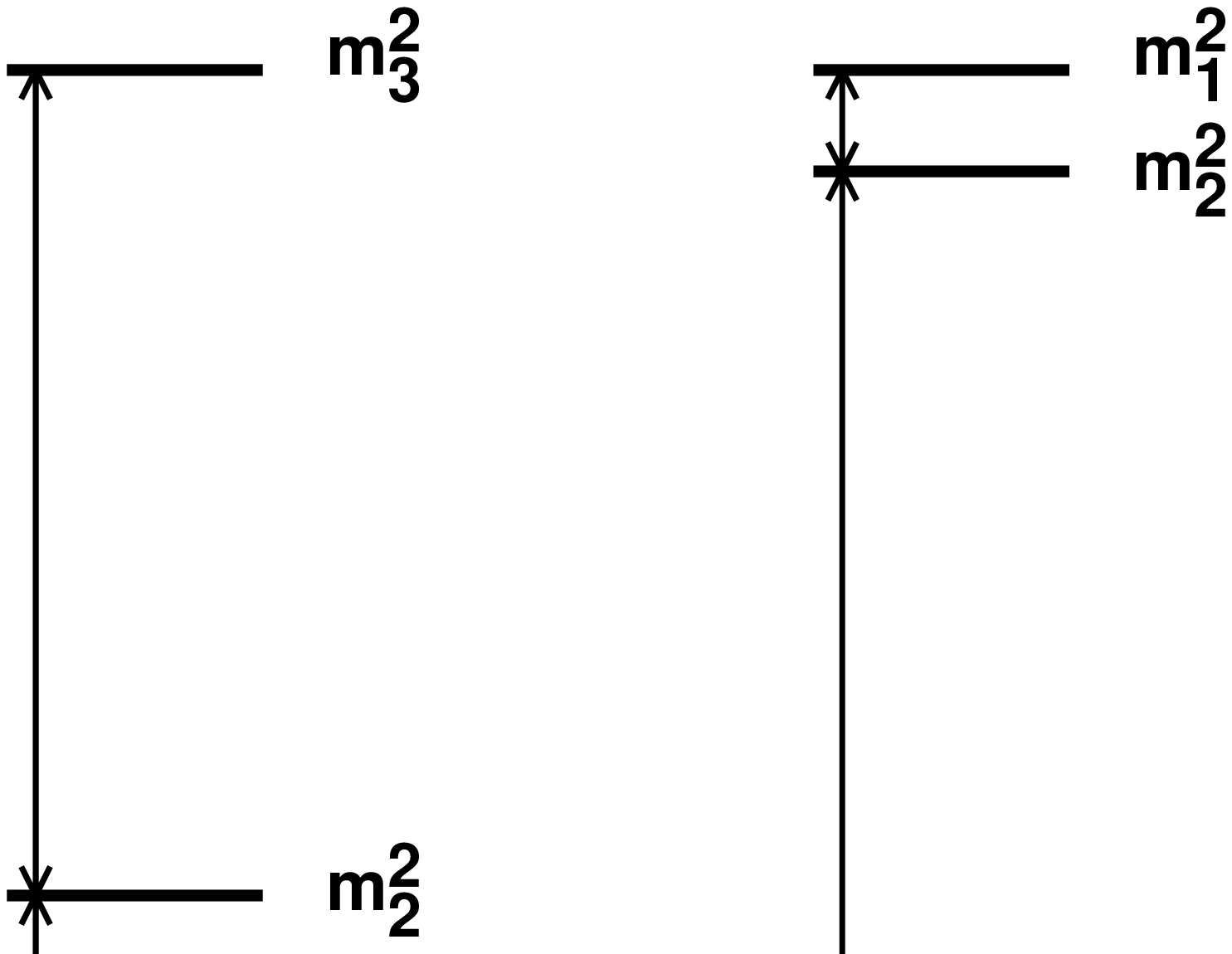,width=7cm}

\vglue 3.5cm
It has been shown in the three flavor framework \cite{yasuda1} that combination
of the CHOOZ reactor data \cite{chooz}
and the atmospheric neutrino data of
the Kamiokande and the Superkamiokande implies very small
$\theta_{13}$, i.e., $\sin^22\theta_{13} < 0.1$ which is
essentially the result of the CHOOZ data.  When $|\theta_{13}|$ is
small, the MNS matrix looks like
\begin{eqnarray}
U\simeq\left(
\begin{array}{lll}
{\ }~c_\odot  & {\ }~s_\odot  &  \epsilon\nonumber\\
-s_\odot c_{\mbox{\rm atm}} & 
{\ }~c_\odot c_{\mbox{\rm atm}} & s_{\mbox{\rm atm}}\nonumber\\
{\ }~s_\odot s_{\mbox{\rm atm}} & 
-c_\odot s_{\mbox{\rm atm}} & c_{\mbox{\rm atm}}\nonumber\\
\end{array}\right),\nonumber
\end{eqnarray}
where $\theta_{12}$, $\theta_{23}$ have been replaced by $\theta_\odot$
and $\theta_{\mbox{\rm atm}}$, respectively.

The measurement of $\theta_\odot\equiv\theta_{12}$ and
$\theta_{\mbox{\rm atm}}\equiv\theta_{23}$ is expected to be greatly
improved in the future experiments on solar and atmospheric neutrinos,
so the remaining problems in the three flavor framework are to
determine (1) the sign of $\Delta m_{32}^2$, (2) the magnitude of
$\theta_{13}$, (3) the magnitude of the CP phase $\delta$.  Recently a lot of
research have been done on neutrino factories, and the three problems
mentioned above may be solved at neutrino factories.
In this talk I would like to discus these three issues in some
detail and show which set of parameters optimizes each signal.
I will assume that the volume of the detector is 10 kt, the
intensity of the beam is 10$^{21}$ muons/yr, and the data are
taken for one year as the reference values in the following
discussions.  I will also assume $E_\mu\le$ 50 GeV.

\section{Neutrino factories}
Before discussing the three problems given at the end of the Introduction
let me give a little background for neutrino factories.
As has been shown in \cite{geer,nf}, the information of neutrino oscillations
can be obtained by looking at ``wrong sign muons'' which are produced
in $\nu_e\rightarrow\nu_\mu\rightarrow\mu^-$ or 
$\bar\nu_e\rightarrow\bar\nu_\mu\rightarrow\mu^+$ and
the numbers $N_{\mbox{\rm wrong}}(\mu^\pm)$ of the wrong sign muons
are given by \cite{geer}
\begin{eqnarray}
N_{\mbox{\rm wrong}}(\mu^-)&=&
n_T{12E_\mu^2 \over \pi L^2 m_\mu^2}
\int d\left({E_\nu \over E_\mu}\right)
\left({E_\nu \over E_\mu}\right)^2
\left(1-{E_\nu \over E_\mu}\right)
\sigma_{\nu N}(E_\nu)P(E_\nu)\nonumber\\
N_{\mbox{\rm wrong}}(\mu^+)&=&
n_T{12E_\mu^2 \over \pi L^2 m_\mu^2}
\int d\left({E_{\bar\nu} \over E_\mu}\right)
\left({E_{\bar\nu} \over E_\mu}\right)^2
\left(1-{E_{\bar\nu} \over E_\mu}\right)
\sigma_{{\bar\nu} N}(E_{\bar\nu})P(E_{\bar\nu}),\nonumber
\end{eqnarray}
where $E_\mu$ is the muon energy, $L$ is the length of the neutrino path,
$n_T$ is the number of the target nucleons,
$\sigma_{\nu N}(E_\nu)$ and
$\sigma_{{\bar\nu} N}(E_{\bar\nu})$ are
the (anti-)neutrino nucleon cross sections given by
\begin{eqnarray}
\sigma_{\nu N}(E_\nu)&=&
\left({E_\nu \over \mbox{\rm GeV}}\right)
\times 0.67 \times 10^{-38}\mbox{\rm cm}^2\nonumber\\
\sigma_{{\bar\nu} N}(E_{\bar\nu})&=&
\left({E_{\bar\nu} \over \mbox{\rm GeV}}\right)
\times 0.33 \times 10^{-38}\mbox{\rm cm}^2,\nonumber
\end{eqnarray}
and $P(E_\nu)$ and $P(E_{\bar\nu})$ are the oscillation probabilities
which are given by (on the assumption of constant density of the matter)
\begin{eqnarray}
P(E_\nu)&\equiv&
P(\nu_e\rightarrow\nu_\mu)=s^2_{23}\sin^22\theta_{13}^{M^{(-)}}
\sin^2\left({B^{(-)}L \over 2}\right)\nonumber\\
P(E_{\bar\nu})&\equiv&
P({\bar\nu}_e\rightarrow{\bar\nu}_\mu)=s^2_{23}\sin^22\theta_{13}^{M^{(+)}}
\sin^2\left({B^{(+)}L \over 2}\right),
\label{eqn:prob}
\end{eqnarray}
where $A\equiv\sqrt{2} G_F N_e$ stands for the matter effect
\cite{msw} of the Earth,
$\theta_{13}^{M^{(\pm)}}$ is the effective mixing angle in
matter given by
\begin{eqnarray}
\tan2\theta_{13}^{M^{(\pm)}}\equiv {\Delta E_{32}\sin2\theta_{13}
\over \Delta E_{32}\cos2\theta_{13}\pm A},
\label{eqn:thetam}
\end{eqnarray}
\noindent
and
\begin{eqnarray}
B^{(\pm)}\equiv \sqrt{\left(\Delta E_{32}\cos2\theta_{13}\pm A\right)^2
+\left(\Delta E_{32}\sin2\theta_{13}\right)^2}.
\label{eqn:b}
\end{eqnarray}
Using these formula, the ratio $N_{\mbox{\rm
wrong}}(\mu^+)/N_{\mbox{\rm correct}}(\mu^-)$ is plotted in Fig. 2a
and 2b as a function of $E_\mu$ and $L$ for typical values of
$\theta_{13}$ with $\sin^22\theta_{23}=1.0$ and
$\Delta m_{32}^2=3.5\times 10^{-3}$eV$^2$.
As was shown by Gomez-Cadenas \cite{gomez} the
ratio of (background events)/$N_{\mbox{\rm correct}}(\mu)$ is of order
10$^{-5}$.  In the case of smaller value of $\theta_{13}$
($\theta_{13}=1^\circ$ or $\sin^22\theta_{13}=0.001$),
it should be possible to detect wrong
sign events at neutrino factories if $L>$1000km for
all the muon energies ($10\le E_\mu\le 50$ GeV), with our reference
values of the beam and the detector
(10$^{21} \mu$/yr$\cdot$10kt$\cdot$1yr).

\vglue -0.7cm \hglue -7.8cm
\epsfig{file=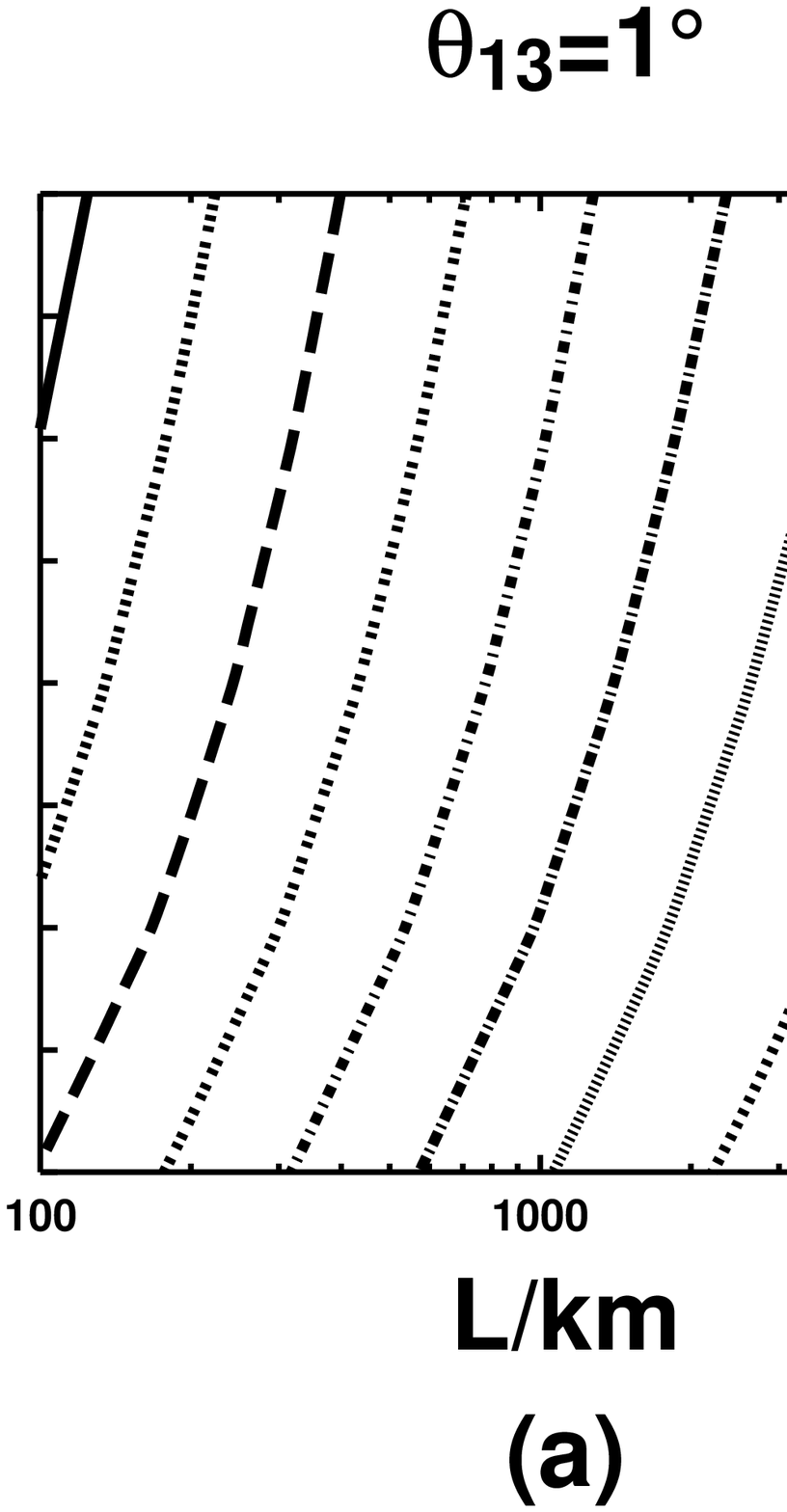,width=9cm}
\vglue -9.1cm \hglue -0.7cm
\epsfig{file=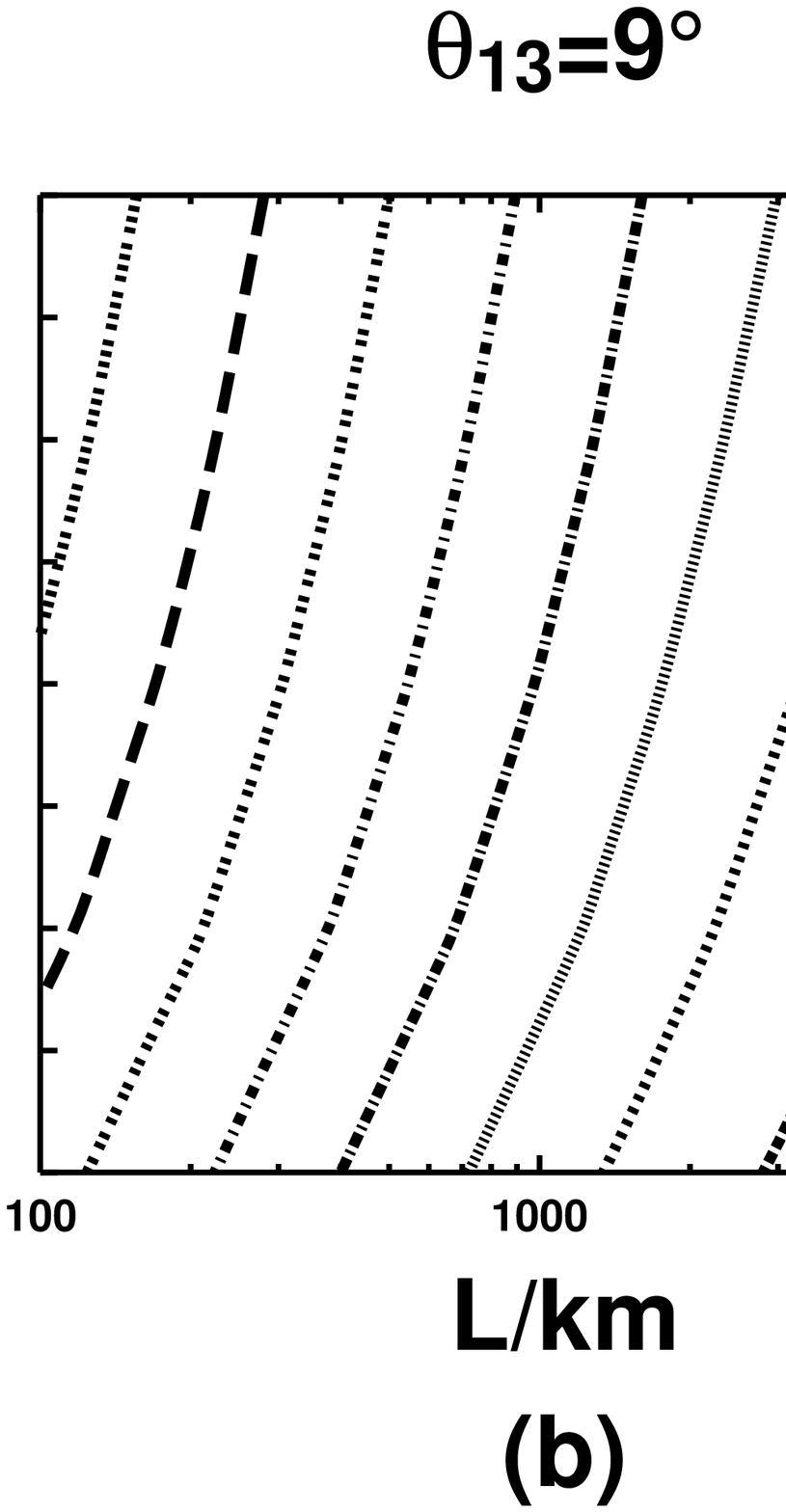,width=9cm}
\vglue -1.6cm \hglue 4.cm
\epsfig{file=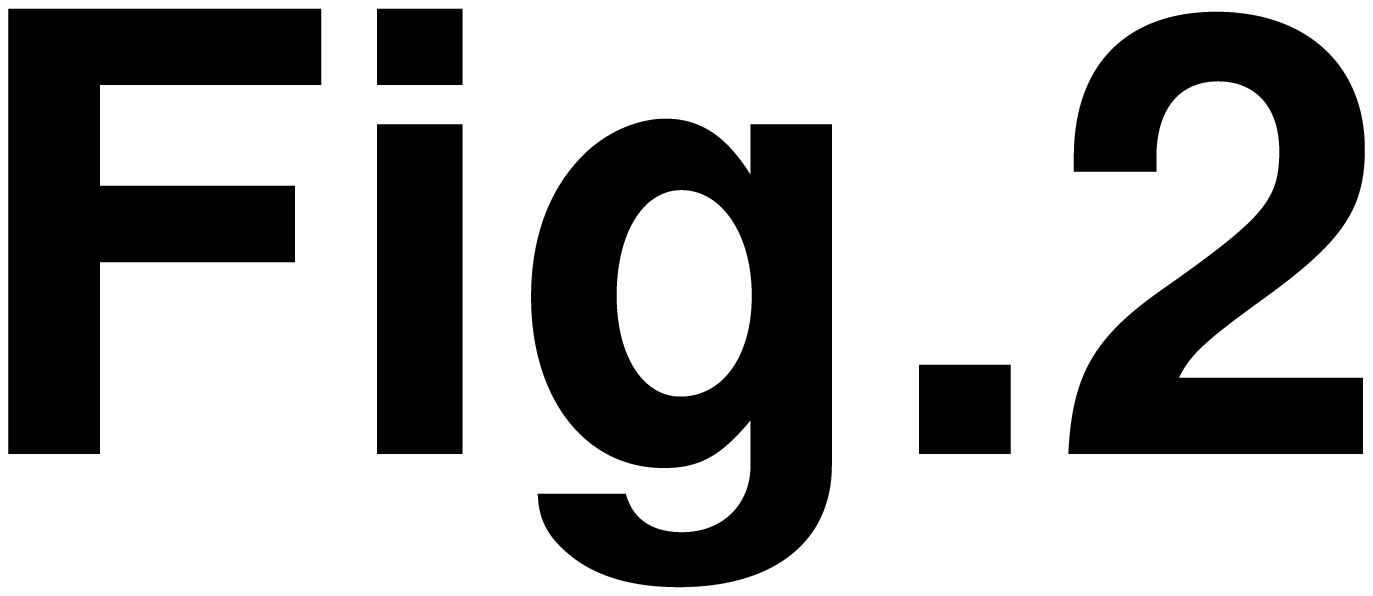,width=2cm}

\vglue -1.5cm
\section{The sign of $\Delta m_{32}^2$}
As was mentioned in the Introduction, the mass pattern corresponds to
either Fig. 1a or 1b, depending on whether $\Delta m_{32}^2$ is
positive or negative.
Determination of this mass pattern is
important, since Figs. 1a and 1b correspond to one and two mass states,
assuming that the lowest mass is almost zero\footnote{The mixed dark
scenario in which neutrinos have masses of order 1 eV seems to be
disfavored \cite{dm}.}.

As we can see from (\ref{eqn:prob}), if $\Delta m_{32}^2>0$
then the effective mixing angle
$\theta_{13}^{M(-)}$ is enhanced and
$P(\nu_e\rightarrow\nu_\mu)$ increases.
On the other hand, if $\Delta m_{32}^2<0$ then 
$\theta_{13}^{M(+)}$ is enhanced and 
$P({\bar\nu}_e\rightarrow{\bar\nu}_\mu)$ increases.
So, at neutrino factories where baseline is relatively large and
therefore the matter effect plays an important role,
the sign of $\Delta m_{32}^2$ may be determined by looking at
the difference between neutrino and anti-neutrino events
which should reflect the difference between 
$P(\nu_e\rightarrow\nu_\mu)$ and
$P({\bar\nu}_e\rightarrow{\bar\nu}_\mu)$.
In Figs. 3 and 4 the numbers of events per each neutrino energy
$E_\nu$ ($E_{\bar\nu}$ and $E_\nu$ is identified here)
at a neutrino factory with the parameter 
10$^{21} \mu$/yr$\cdot$10kt$\cdot$1yr are given for
$\Delta m_{32}^2=3.5\times 10^{-3}$eV$^2$ (Fig. 3) and
$\Delta m_{32}^2=-3.5\times 10^{-3}$eV$^2$ (Fig. 4),
respectively (I have assumed $\sin^22\theta_{23}=1.0$ and
$\sin^22\theta_{13}=0.095$).
The black and gray lines stand for
wrong sign $\mu^-$ and $\mu^+$ events, respectively,
and five cases of the muon energy
$E_\mu = 10,20,\cdots,50$ GeV are plotted.  For later purposes,
behaviors with respect to the CP violating phase $\delta$ are
also considered and the solid, dotted and dashed lines are the numbers
of events with $\delta=0$, $\delta=-\pi/2$, $\delta=\pi/2$, respectively.
The deviation of the solid lines from the dotted or dashed lines
is mainly due to the matter effects, and the deviation of the
dotted lines from the dashed ones is in general smaller.

Since the cross section $\sigma_{\nu N}$ and $\sigma_{{\bar\nu} N}$
are different (the ratio is 2 to 1), it is useful to look at the
quantity
\begin{eqnarray}
{N_\nu-2N_{\bar\nu} \over \delta(N_\nu-2N_{\bar\nu})}
={N_\nu-2N_{\bar\nu} \over \sqrt{N_\nu+4N_{\bar\nu}}}\nonumber
\end{eqnarray}
whose absolute value should be much larger than one
to demonstrate $\Delta m_{32}^2>0$ or $\Delta m_{32}^2<0$.
Now let me introduce the quantity
\begin{eqnarray}
R\equiv 
{\left[N_{\mbox{\rm wrong}}(\mu^-)-2N_{\mbox{\rm wrong}}(\mu^+)\right]^2
\over N_{\mbox{\rm wrong}}(\mu^-)+4N_{\mbox{\rm wrong}}(\mu^+)}.
\nonumber
\end{eqnarray}
If $R\gg 1$ then we can deduce the sign of
$\Delta m_{32}^2$.
The contour plot of $R$=const. is given in Fig. 5a and 5b
for typical values of $\theta_{13}$ with
$\Delta m_{32}^2=3.5\times 10^{-3}$eV$^2>$0,
$\sin^22\theta_{23}=1.0$.
If $\sin^22\theta_{13}$ is not
smaller than $10^{-3}$, it is possible to determine the sign of
$\Delta m_{32}^2$.  Irrespective of the value of $\theta_{13}$,
$L\sim$ 5000km, $E_\mu=$ 50 GeV seem to optimize the signal,
as far as the quantity $R$ is concerned.

\newpage
\vglue -4.4cm \hglue -5.cm
\epsfig{file=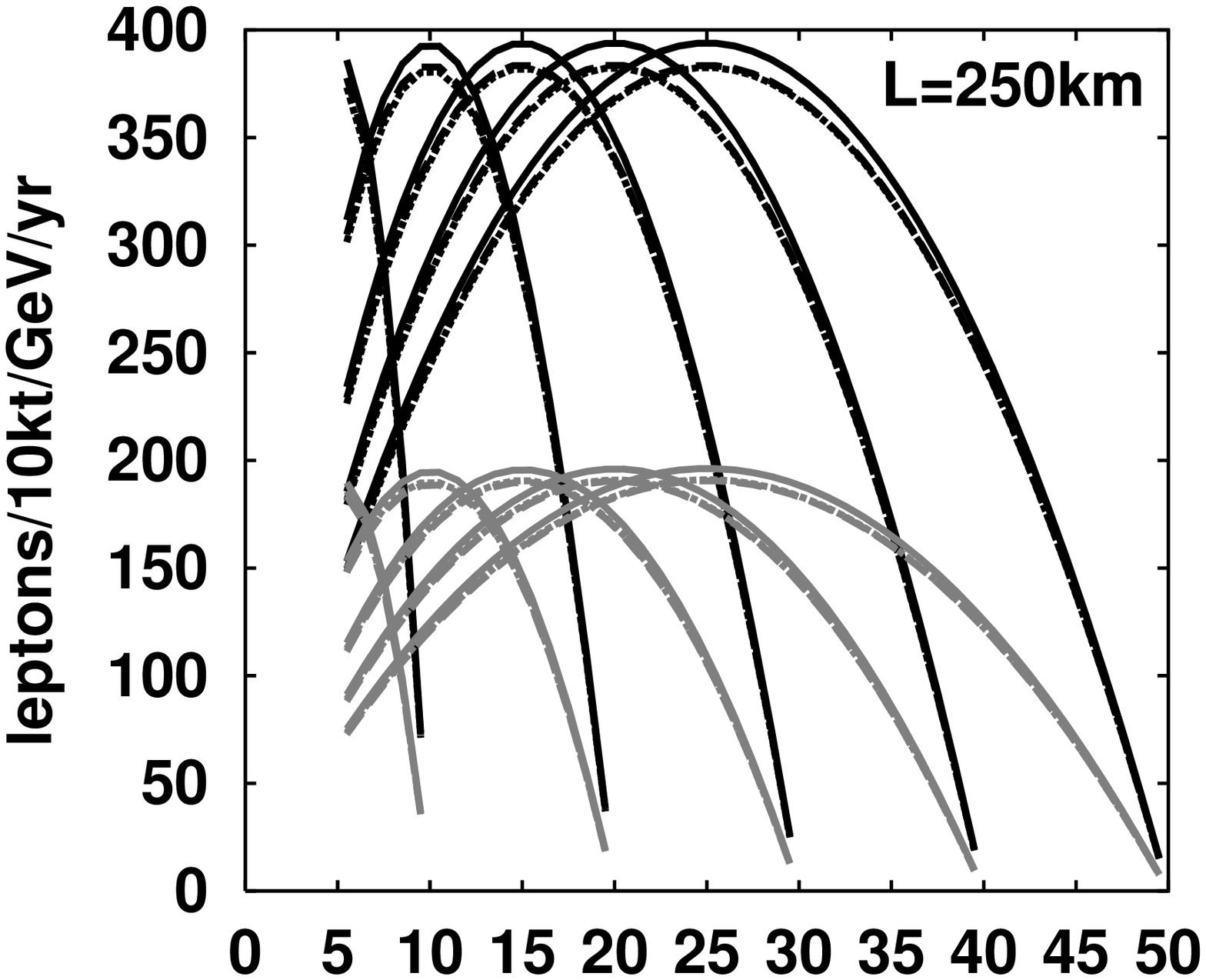,width=10cm}
\vglue -10.cm \hglue -0.4cm
\epsfig{file=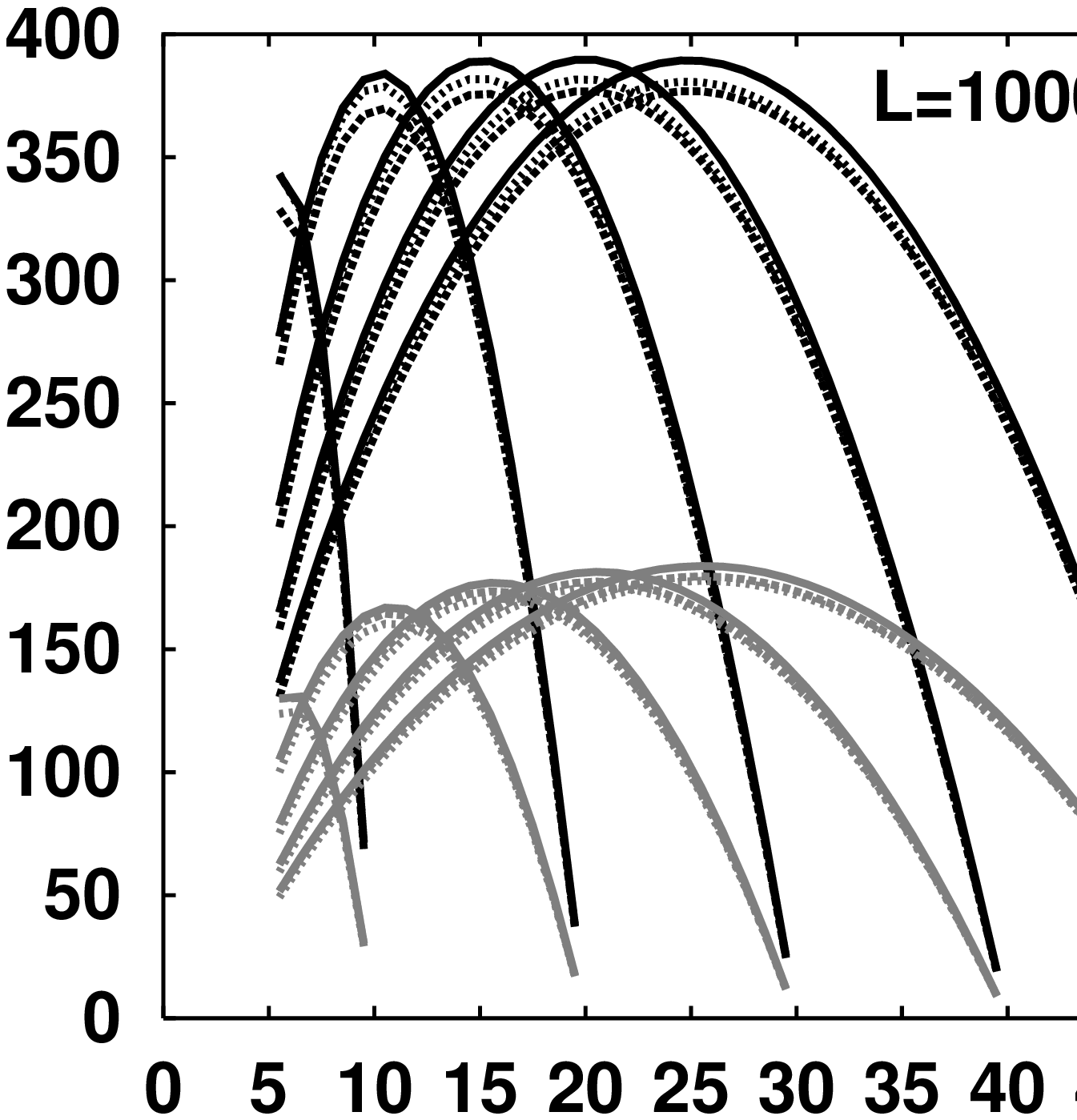,width=10cm}
\vglue -4.5cm \hglue -6.7cm
\epsfig{file=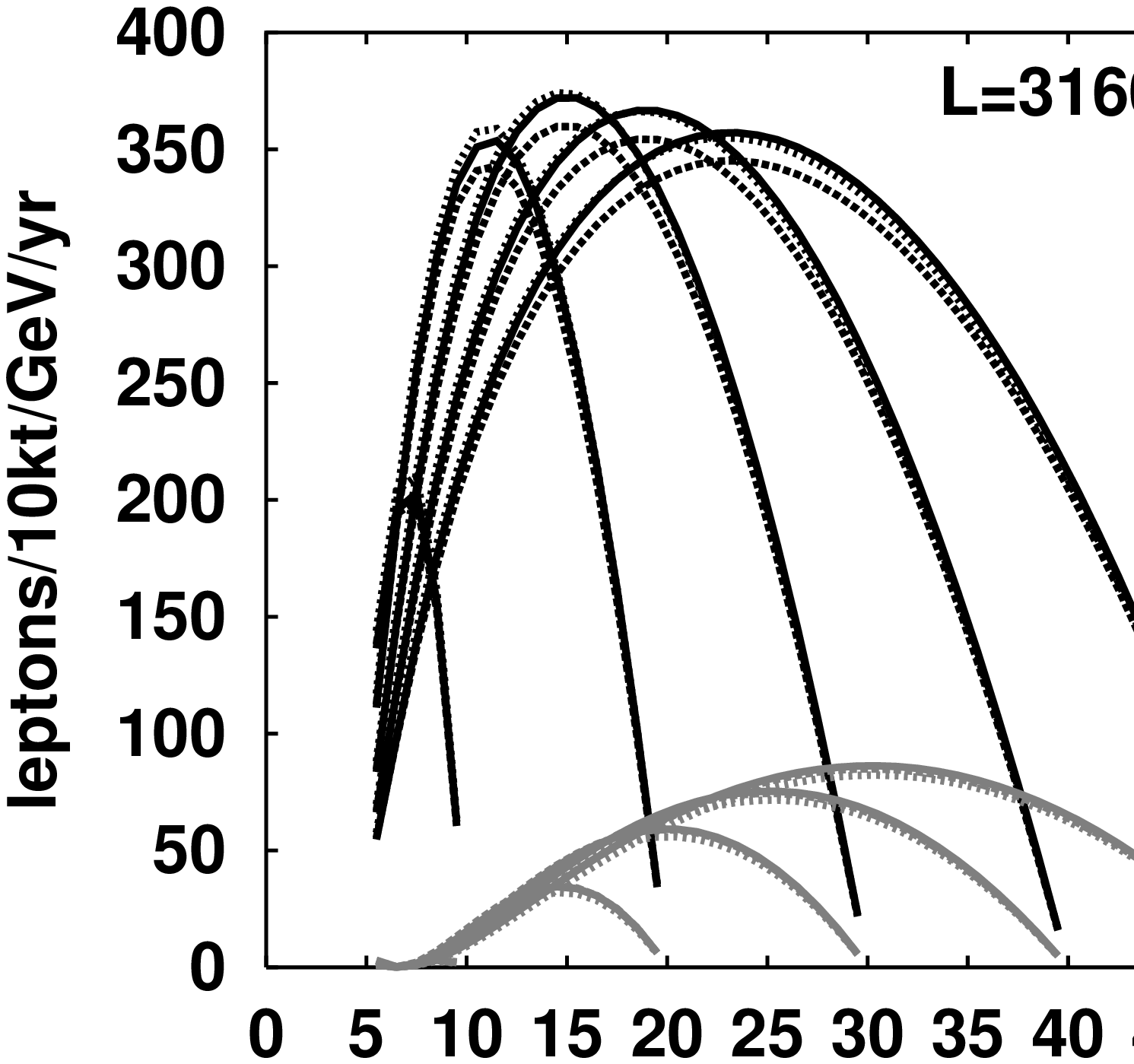,width=10cm}
\vglue -10.cm \hglue -0.4cm
\epsfig{file=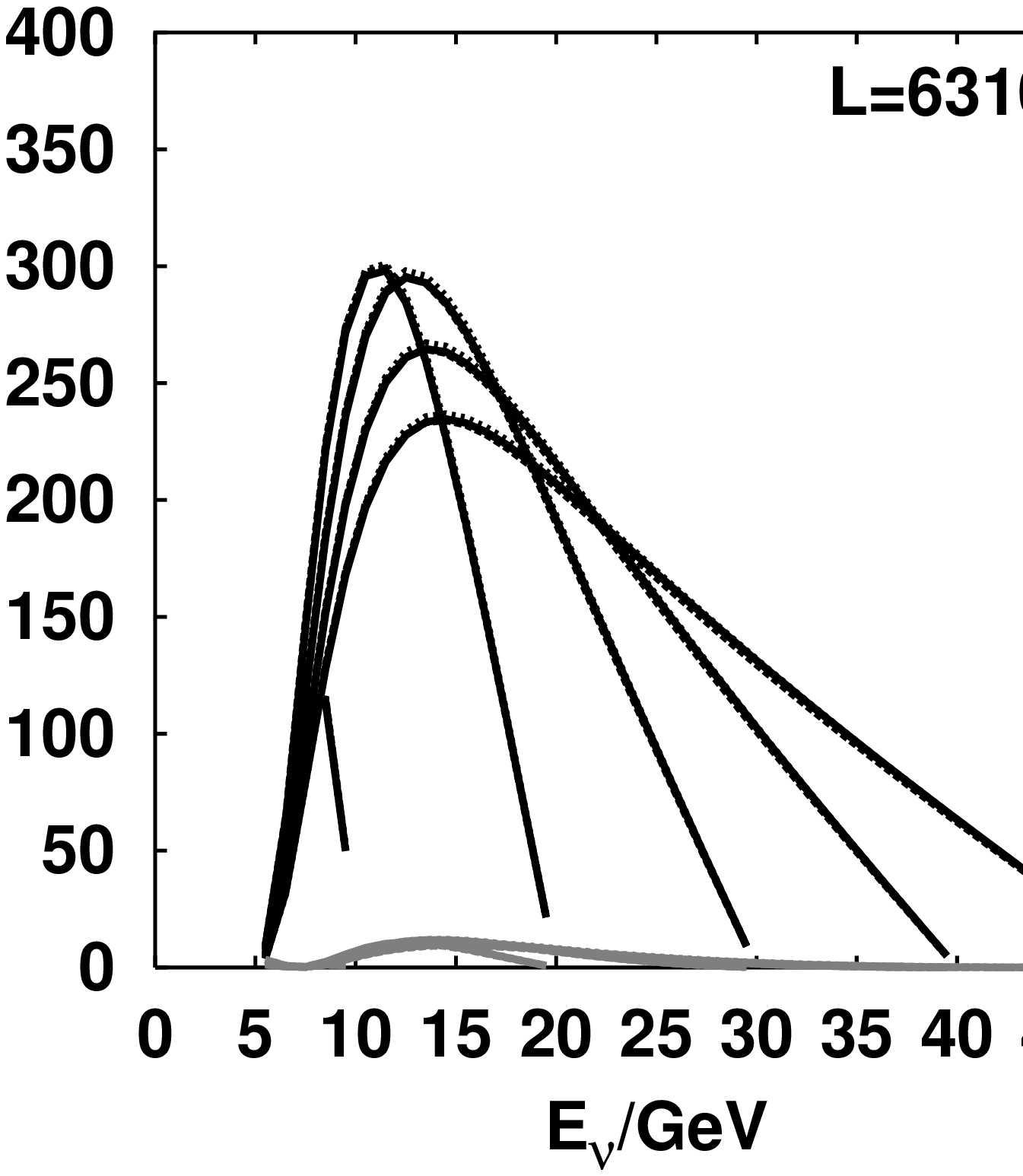,width=10cm}
\vglue -4.5cm \hglue -6.7cm
\epsfig{file=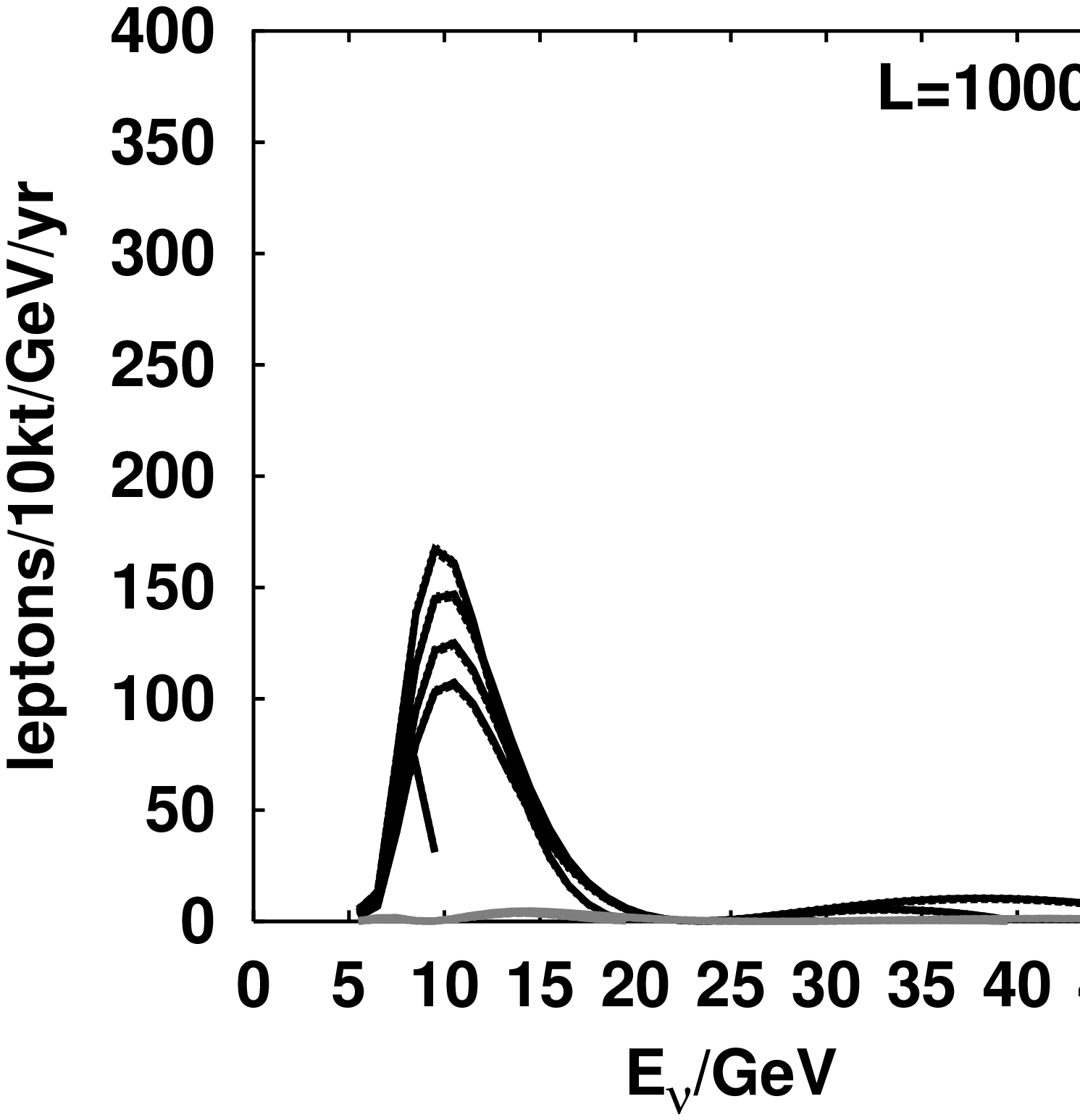,width=10cm}
\vglue -10.5cm \hglue -8.5cm
\epsfig{file=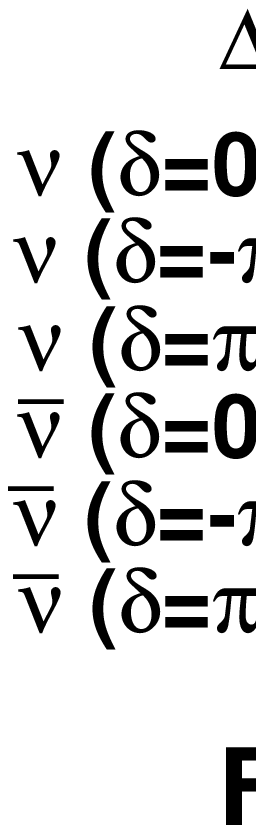,width=15cm}

\newpage
\vglue -4.4cm \hglue -5.cm
\epsfig{file=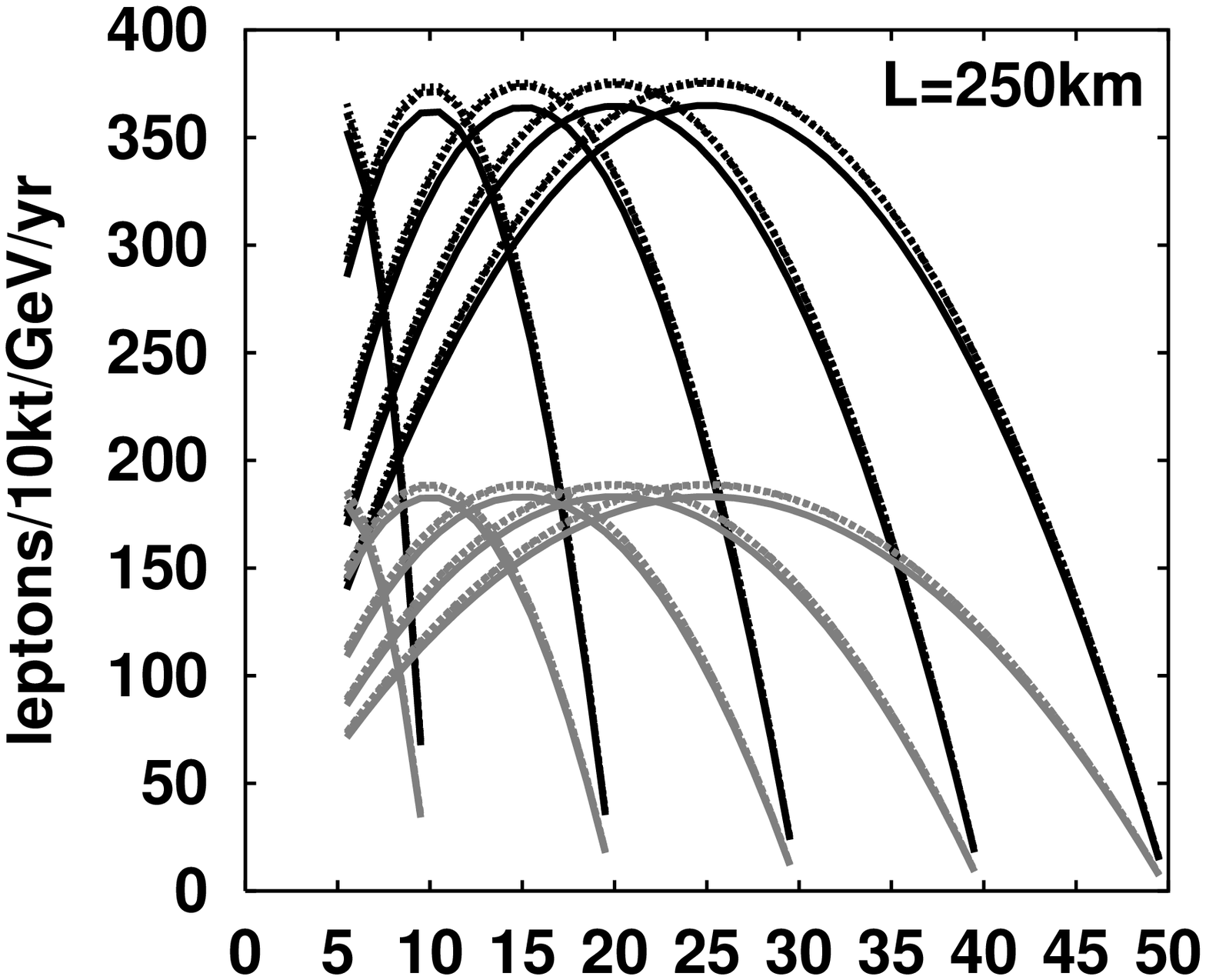,width=10cm}
\vglue -10.cm \hglue 1.2cm
\epsfig{file=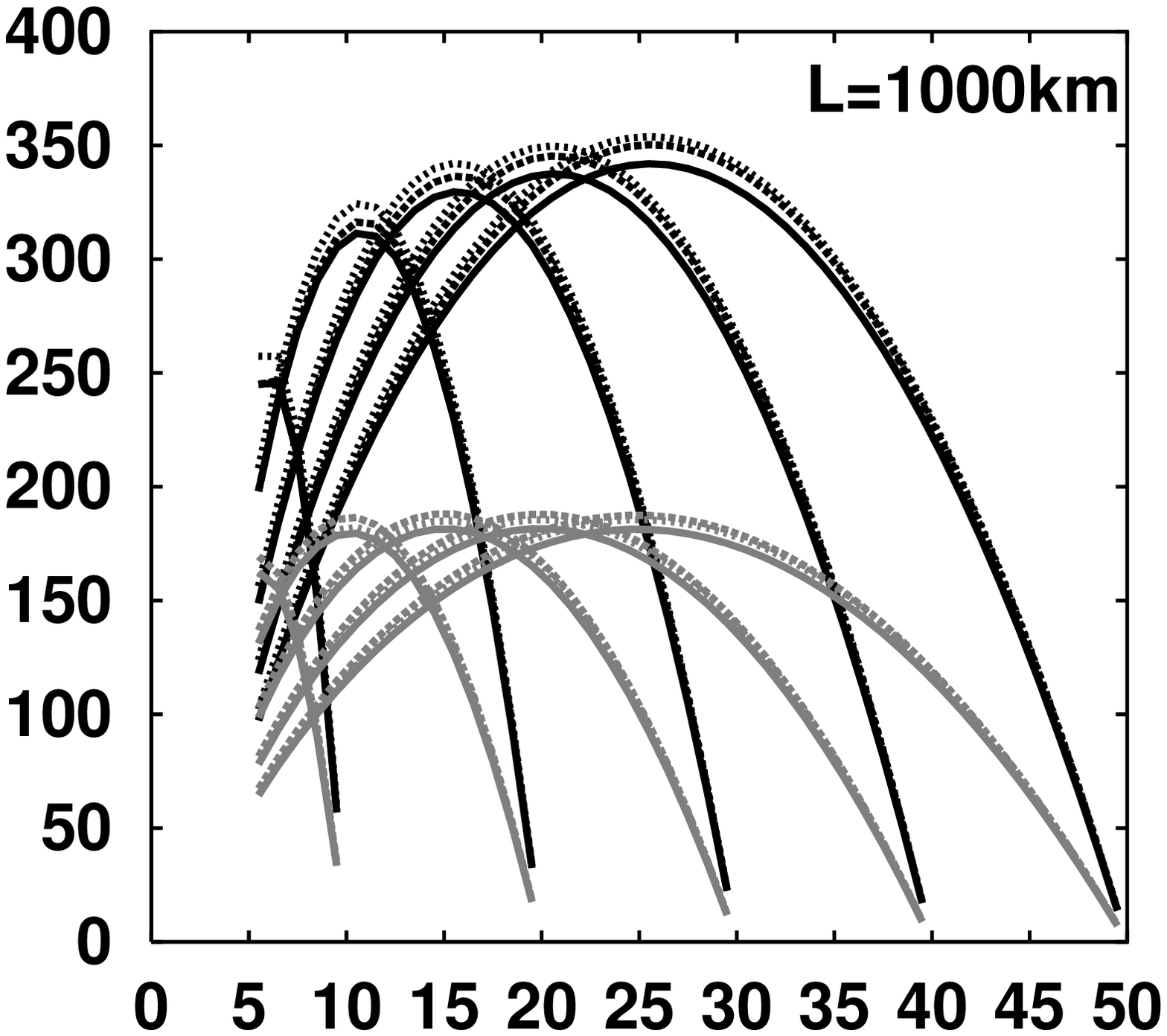,width=10cm}
\vglue -4.5cm \hglue -5.cm
\epsfig{file=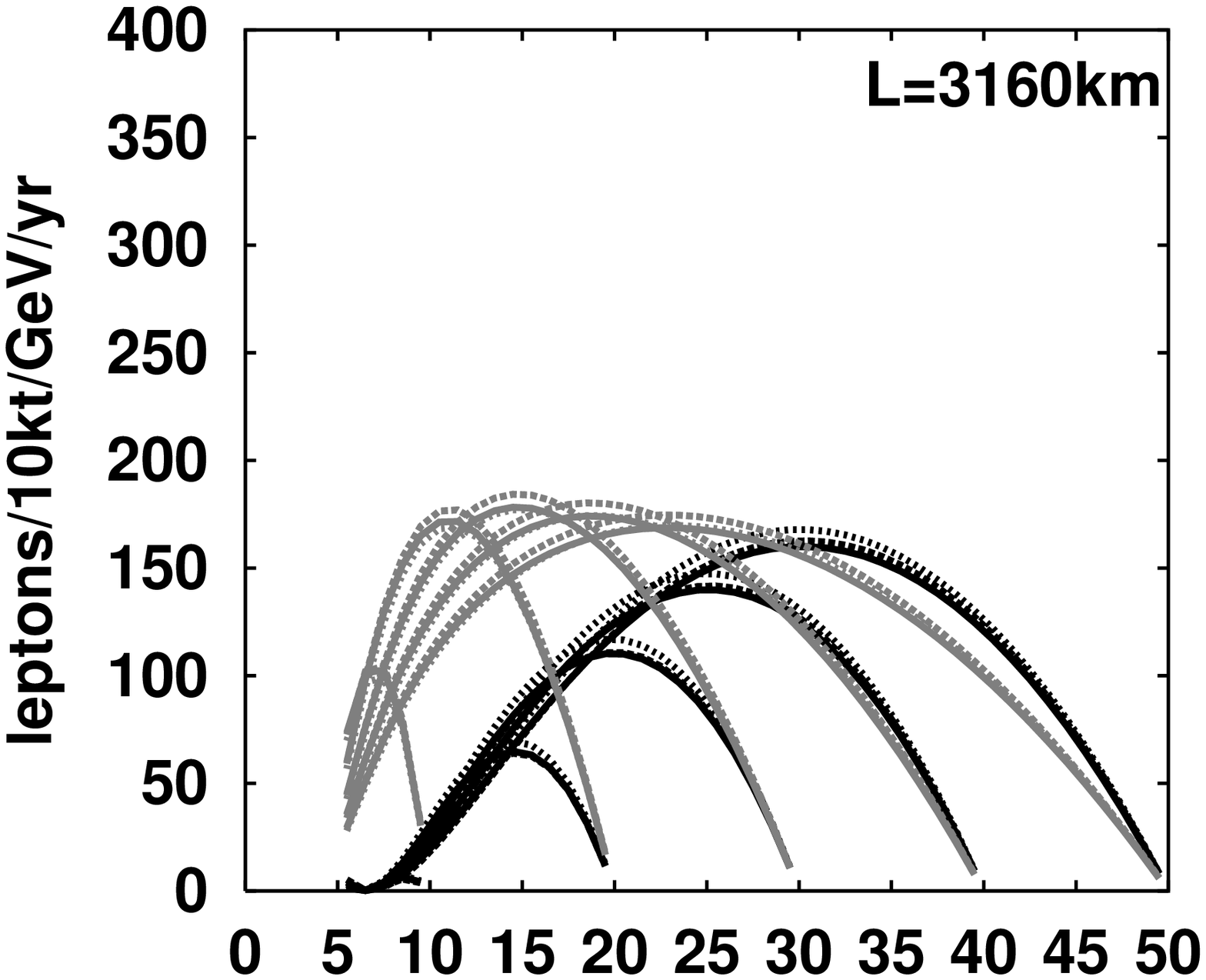,width=10cm}
\vglue -10.cm \hglue 1.2cm
\epsfig{file=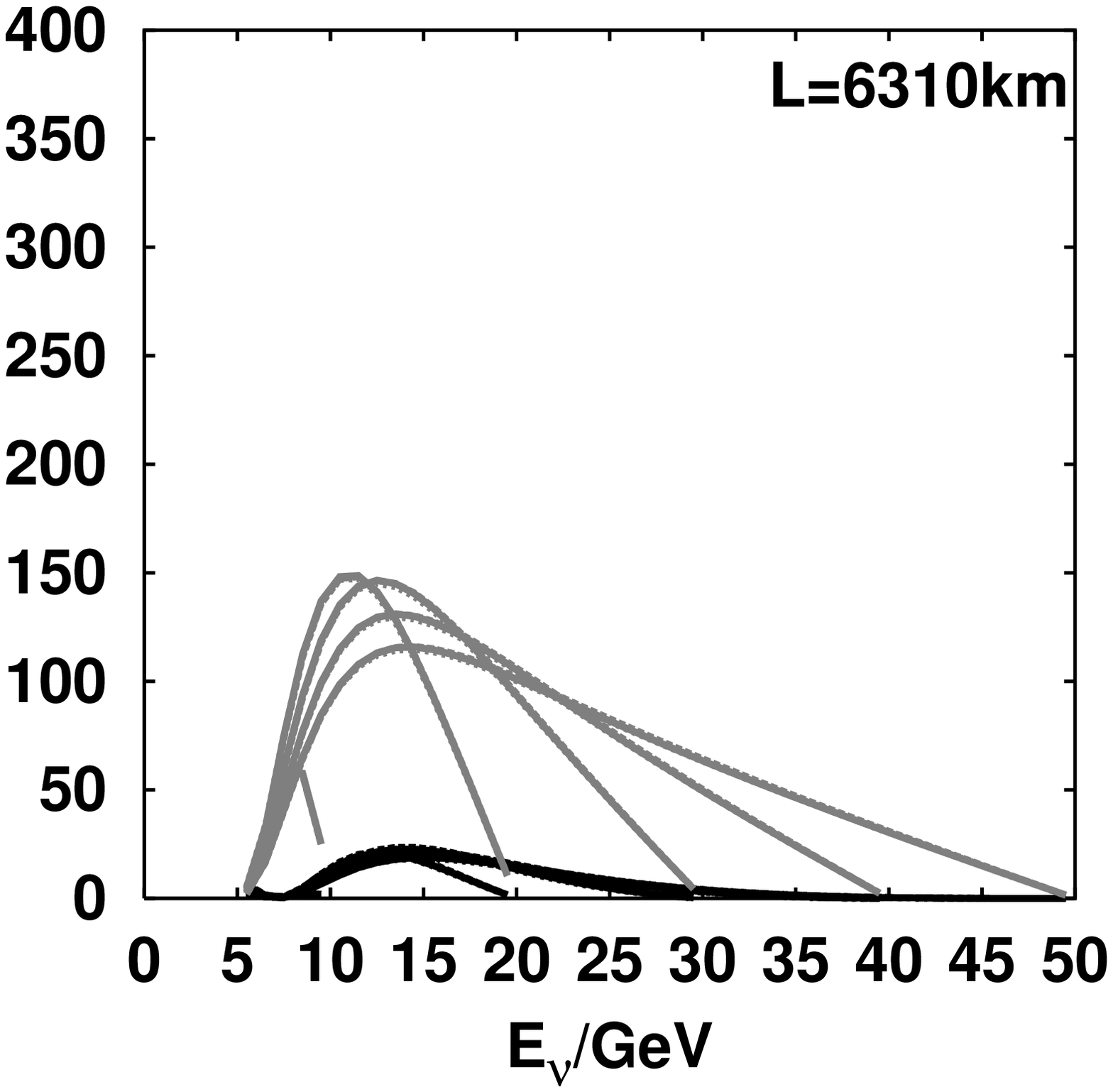,width=10cm}
\vglue -4.5cm \hglue -6.6cm
\epsfig{file=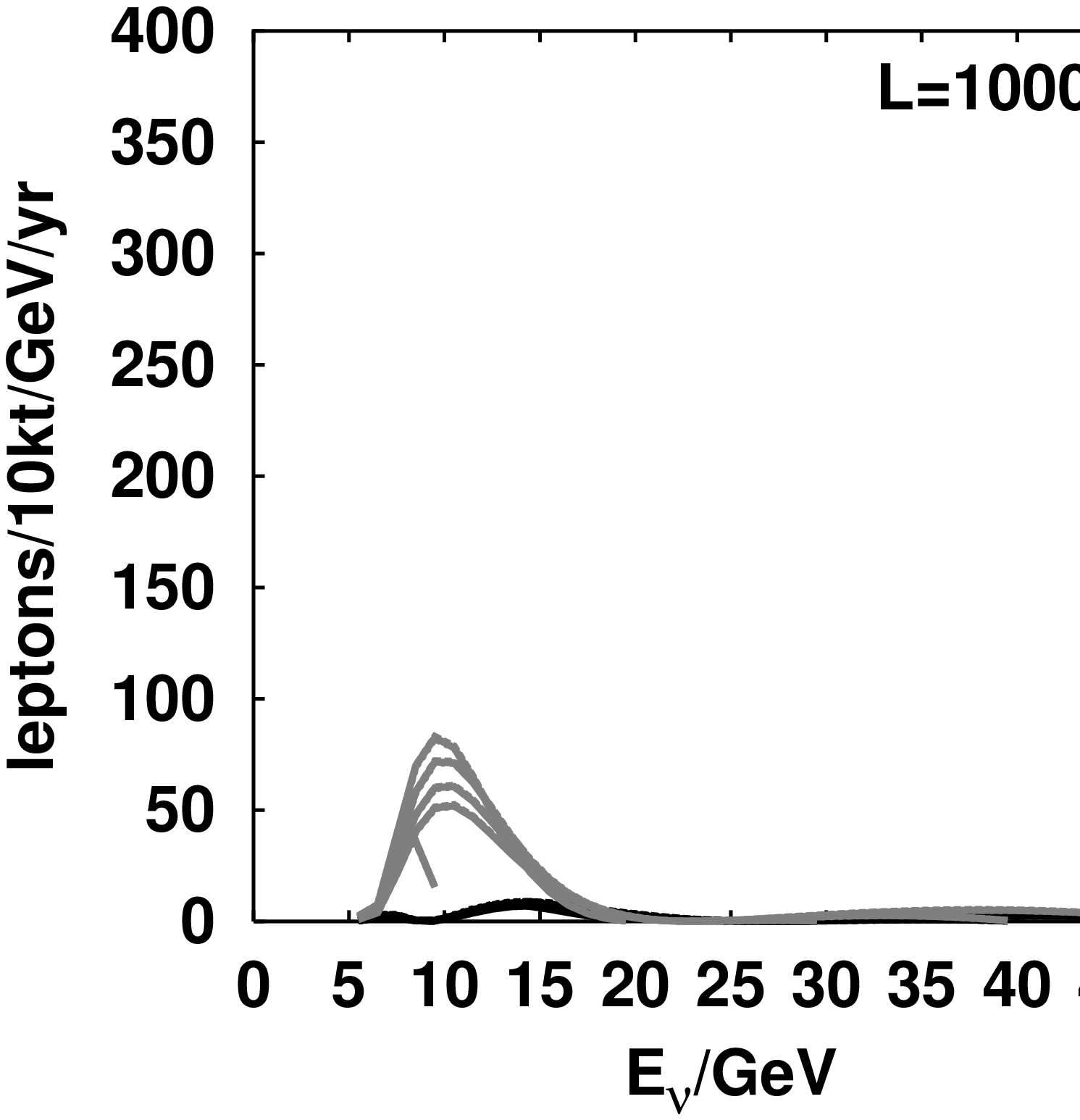,width=10cm}
\vglue -10.5cm \hglue -8.5cm
\epsfig{file=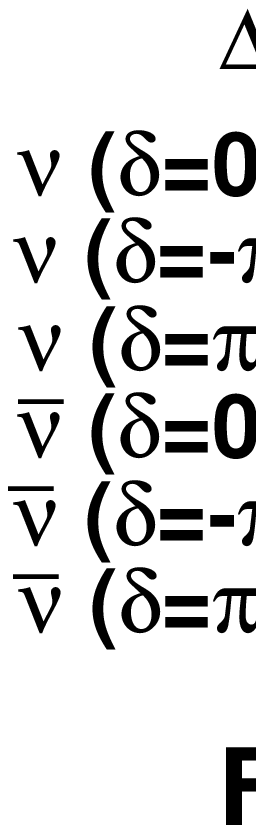,width=15cm}
\newpage
\vglue -0.7cm \hglue -7.8cm
\epsfig{file=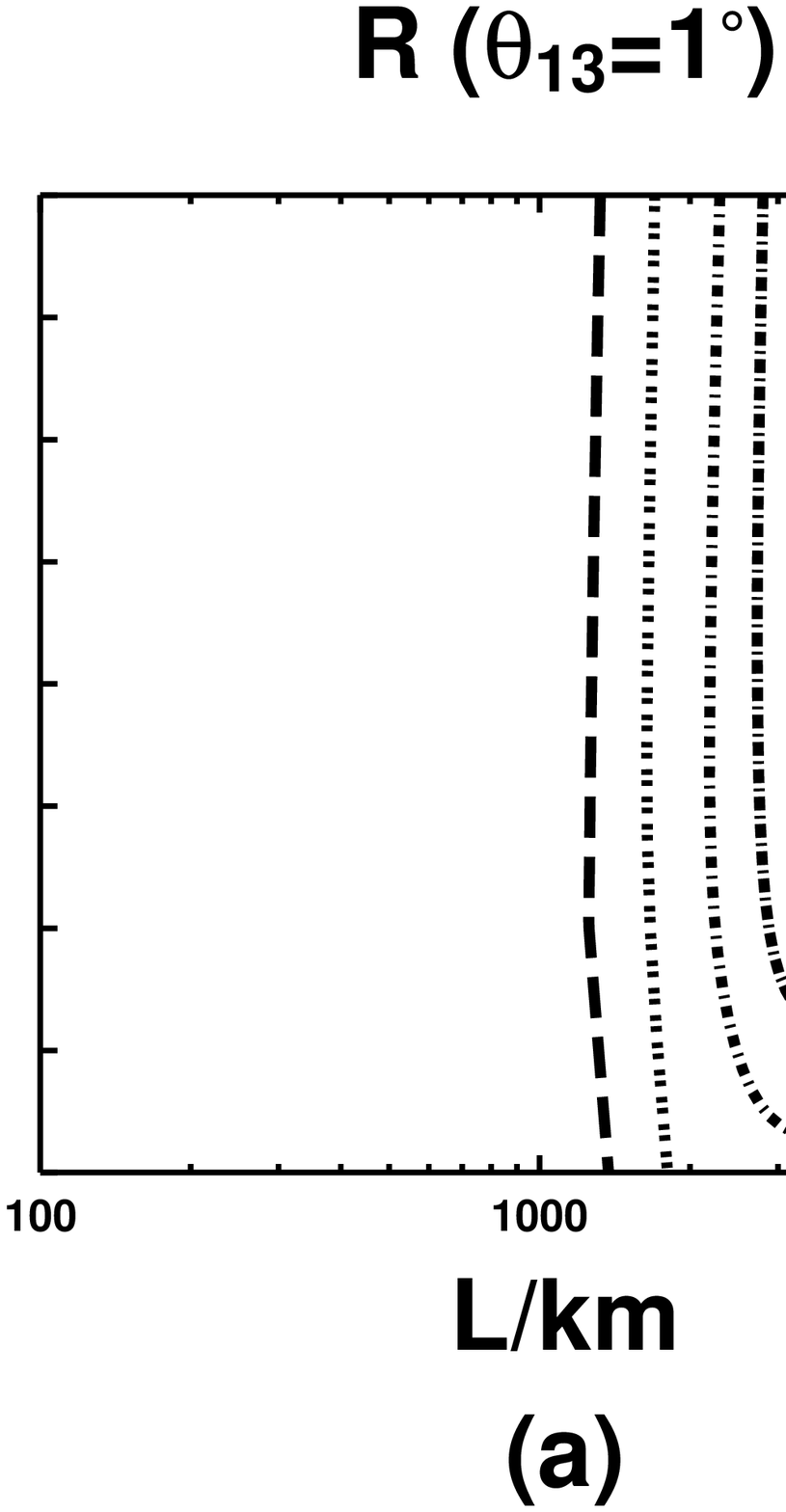,width=9cm}
\vglue -9.1cm \hglue -0.7cm
\epsfig{file=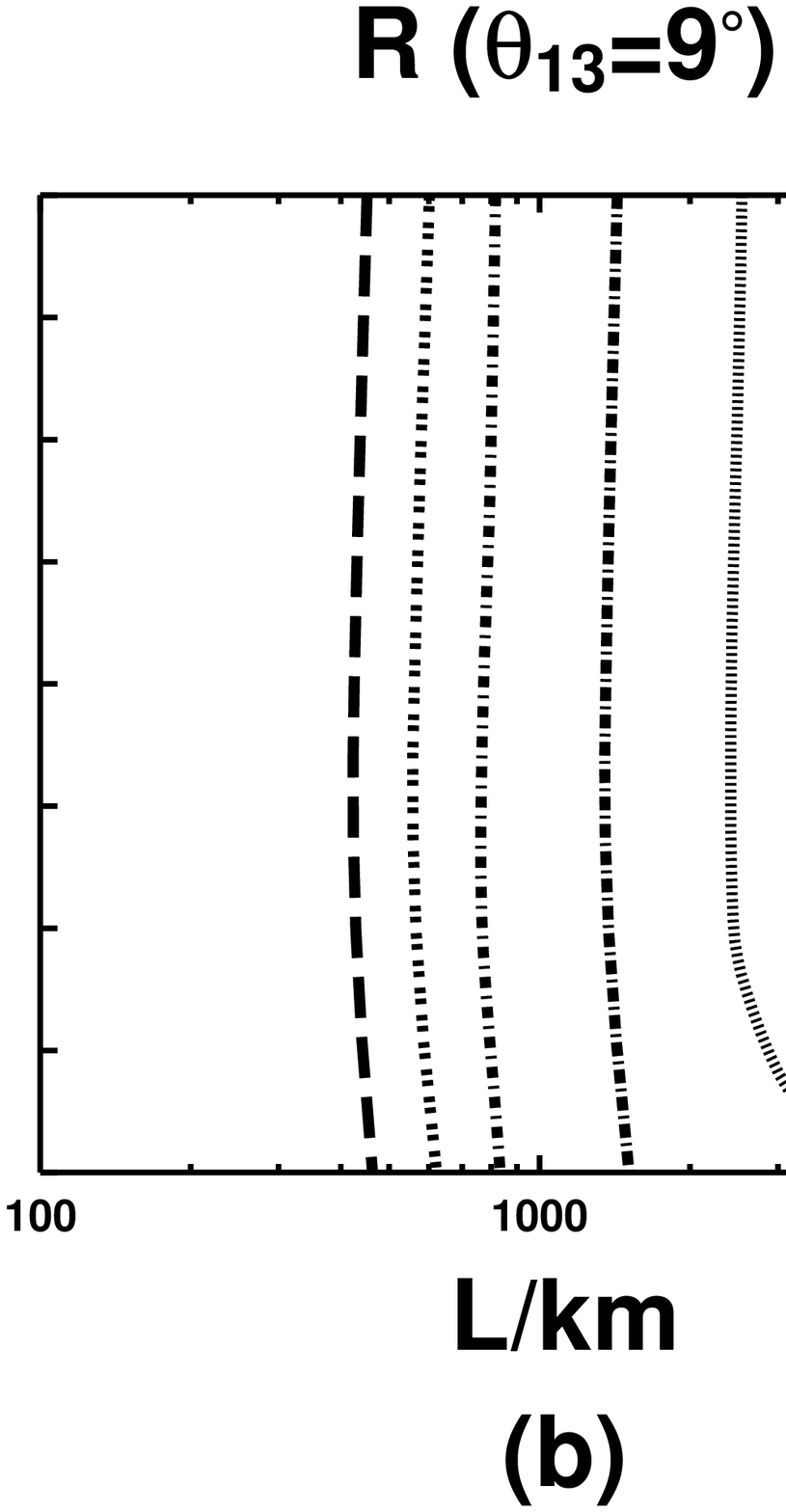,width=9cm}
\vglue -1.6cm \hglue 4.cm
\epsfig{file=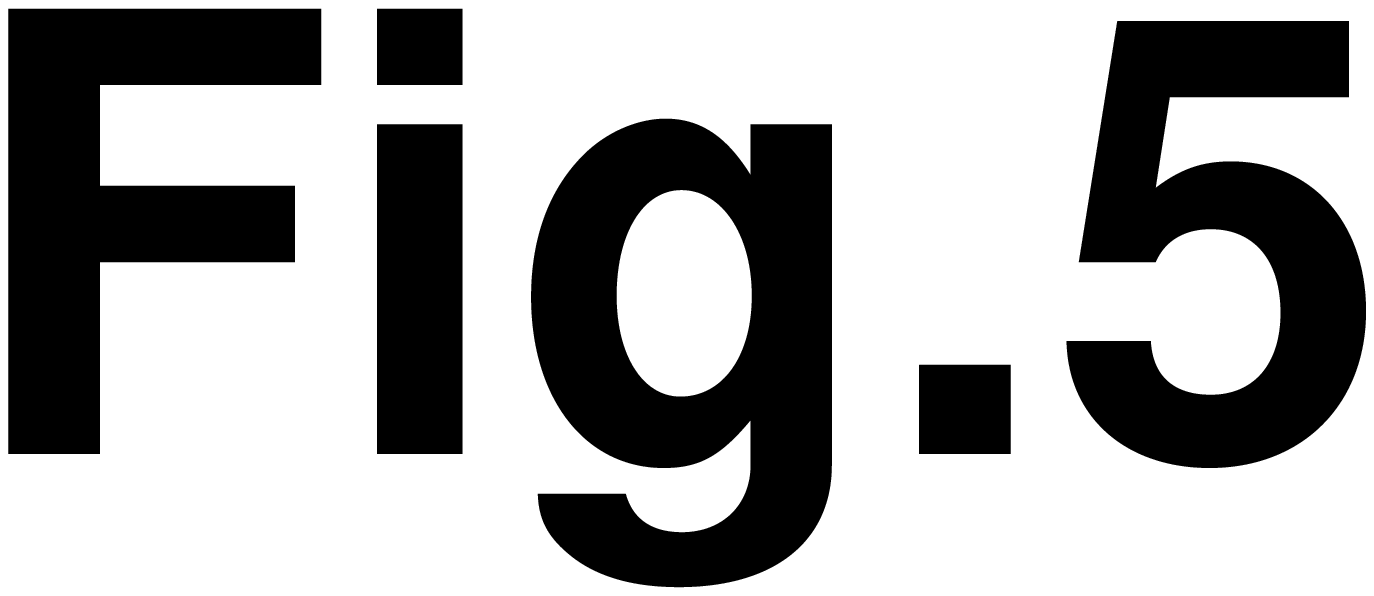,width=2cm}

\section{The magnitude of $\theta_{13}$}
As we can see in (\ref{eqn:prob}), it is necessary to know
the precise value of $\theta_{23}$ to determine $\theta_{13}$
accurately.  For this purpose, it is useful to measure
the independent quantity
\begin{eqnarray}
1-P(\nu_\mu\rightarrow\nu_\mu)&=&
s^4_{23}\sin^22\theta_{13}^{M^{(-)}}
\sin^2\left({B^{(-)}L \over 2}\right)\nonumber\\
&+&\sin^22\theta_{23}\left[\sin^22\theta_{13}^{M^{(-)}}
\sin^2{L \over 4}\left(\Delta E_{32}+A-B^{(-)}\right)\right.
\nonumber\\
&+&\cos^22\theta_{13}^{M^{(-)}}
\sin^2{L \over 4}\left(\Delta E_{32}+A+B^{(-)}\right)
\left.\right].
\label{eqn:prob2}
\end{eqnarray}

Combining (\ref{eqn:prob}) and (\ref{eqn:prob2}) and assuming that our
knowledge on the density profile of the Earth is exact, we can
determine $\theta_{13}$ and $\theta_{23}$.  In practice, however,
there is always uncertainty in the density in the Earth, particularly
the density deep inside of the Earth (i.e., large $L$) is not very
well known, so to determine $\theta_{13}$ precisely the neutrino path
$L$ had better be small, say, $L<1000$km, as long as $N_{\mbox{\rm
wrong}} (\mu)$ exceeds the number of the background events.

\section{The magnitude of $\delta$}
In the hierarchical limit, to first order in $\Delta E_{21}/
\Delta E_{32}$ and $\Delta E_{21}/A$, the appearance probabilities
are given by
\begin{eqnarray}
\left\{
\begin{array}{c}
P(\nu_e\rightarrow\nu_\mu)\nonumber\\
P({\bar\nu}_e\rightarrow{\bar\nu}_\mu)
\end{array}\right\}
&\simeq& s^2_{23}\sin^22\theta_{13}^{M^{(\mp)}}
\sin^2\left({B^{(\mp)}L \over 2}\right)\nonumber\\
&\mp&{1 \over 2}
{\Delta E_{21}\Delta E_{32} \over \lambda_+^{(\mp)} \lambda_-^{(\mp)}}
\sin\delta\sin2\theta_{12}\sin2\theta_{23}\sin2\theta_{13}^{M^{(\mp)}}
\nonumber\\
&\times&\sin\left({\lambda_+^{(\mp)}L \over 2}\right)
\sin\left({\lambda_-^{(\mp)}L \over 2}\right)
\sin\left({B^{(\mp)}L \over 2}\right),\nonumber
\end{eqnarray}
where
\begin{eqnarray}
\lambda_\pm^{(-)}\equiv {1 \over 2}\left(A-\Delta E_{32} \pm B^{(-)}\right),
~\lambda_\pm^{(+)}\equiv {1 \over 2}\left(A-\Delta E_{32} \pm B^{(+)}\right).
\nonumber
\end{eqnarray}
Because of the matter effect, one of the two independent
probabilities is enhanced while the other is suppressed.
For the enhanced channel ($\nu_e\rightarrow\nu_\mu$ in the case of
$\Delta m_{32}^2>0$, ${\bar\nu}_e\rightarrow{\bar\nu}_\mu$
in the case of $\Delta m_{32}^2<0$), the number of events is large
and one might want to take advantage of this large number.
Here I therefore would like to consider the following
quantity
\begin{eqnarray}
R_\delta\equiv
{\left[N_{\mbox{\rm wrong}}(\delta={\pi \over 2})-N_{\mbox{\rm wrong}}
(\delta=-{\pi \over 2})\right]^2
\over 2\left[N_{\mbox{\rm wrong}}(\delta={\pi \over 2})
+N_{\mbox{\rm wrong}}(\delta=-{\pi \over 2})\right]}.
\label{eqn:rd}
\end{eqnarray}
The denominator in (\ref{eqn:rd}) corresponds to square of statistical
fluctuations of the number of events obtained from the T-invariant
probability
$P(\nu_e\rightarrow\nu_\mu;\delta)+P(\nu_e\rightarrow\nu_\mu;-\delta)$
while the numerator does to squqare of the number of events obtained
from T-violating probability
$P(\nu_e\rightarrow\nu_\mu;\delta)-P(\nu_e\rightarrow\nu_\mu;-\delta)$.
This $R_\delta$ is the quantity of T-violation instead of
CP-violation.  In fact $R_\delta$ cannot be determined by the
experimental data only, but it requires knowledge on $\Delta
m^2_{ij}$, $\theta_{ij}$ and the density profile of the Earth to
deduce $R_\delta$.  Nevertheless, having large value of $R_\delta$ is
a necessary condition to be able to measure $\delta$ and the
experiments should be designed so that $R_\delta$ be maximized.

This suggestion is different from the one in \cite{cp}
in which it is proposed to subtract
$(N_\nu-2N_{\bar\nu})/(N_\nu+2N_{\bar\nu})$ by the matter effect
term.  In either case, one needs the precise knowledge on $\Delta
m^2_{ij}$, $\theta_{ij}$ and the density
of the Earth.
To demonstrate $\delta\ne 0$ it is necessary that $R_\delta\gg1$.
The contour plot of the ratio $R_\delta$ is given in Figs. 6a and 6b
for two sets of parameters.
For the set of the oscillation parameters
$(\Delta m^2_{21},\sin^22\theta_{12})=(1.8\times 10^{-5}
{\rm eV}^2, 0.76)$, $(\Delta m_{32}^2,~\sin^22\theta_{23})
= (3.5\times 10^{-3}{\rm eV}^2,1.0)$ (Fig. 6a), which gives the best fit to the
data of solar and atmospheric neutrinos, we have\footnote{
For solar neutrinos, there are three possible sets of parameters
which give a very good fit to the data, but here I take the
most optimistic one (large mixing angle MSW solution) to
observe CP violation.  For other solar neutrino solutions,
observation of CP violation is either very difficult or impossible.}
\begin{eqnarray}
\max_{L,E_\mu} R_\delta \simeq 2.3
\label{eqn:r}
\end{eqnarray}
for $\sin^22\theta_{13}=0.09$, $L\simeq 3000$km, $\delta=\pi/2$,
$E_\mu\simeq$40 GeV with 10$^{21} \mu$/yr$\cdot$10kt$\cdot$1yr.

\vglue -0.10cm \hglue -6.1cm
\epsfig{file=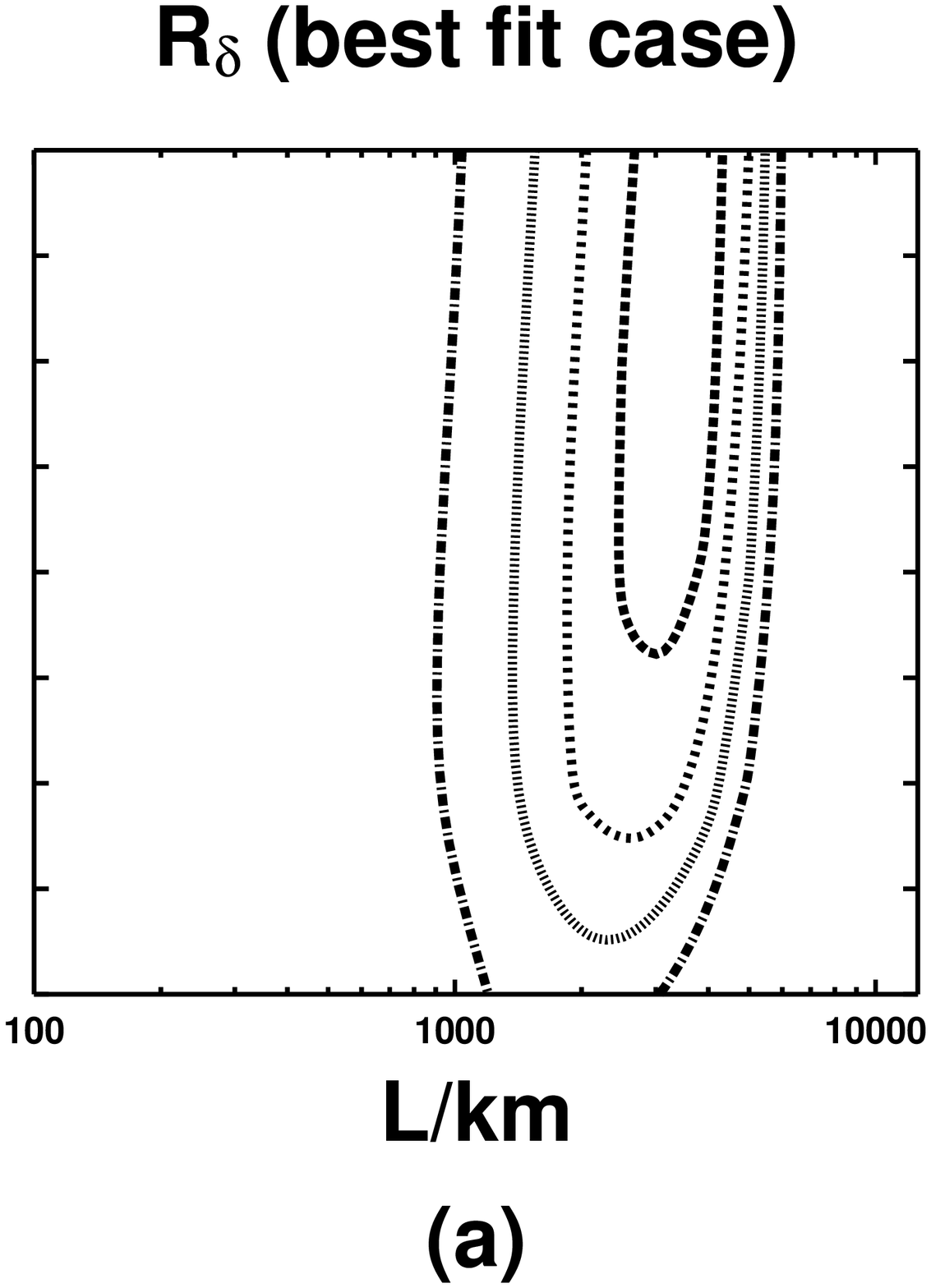,width=9cm}
\vglue -9.1cm \hglue -0.7cm
\epsfig{file=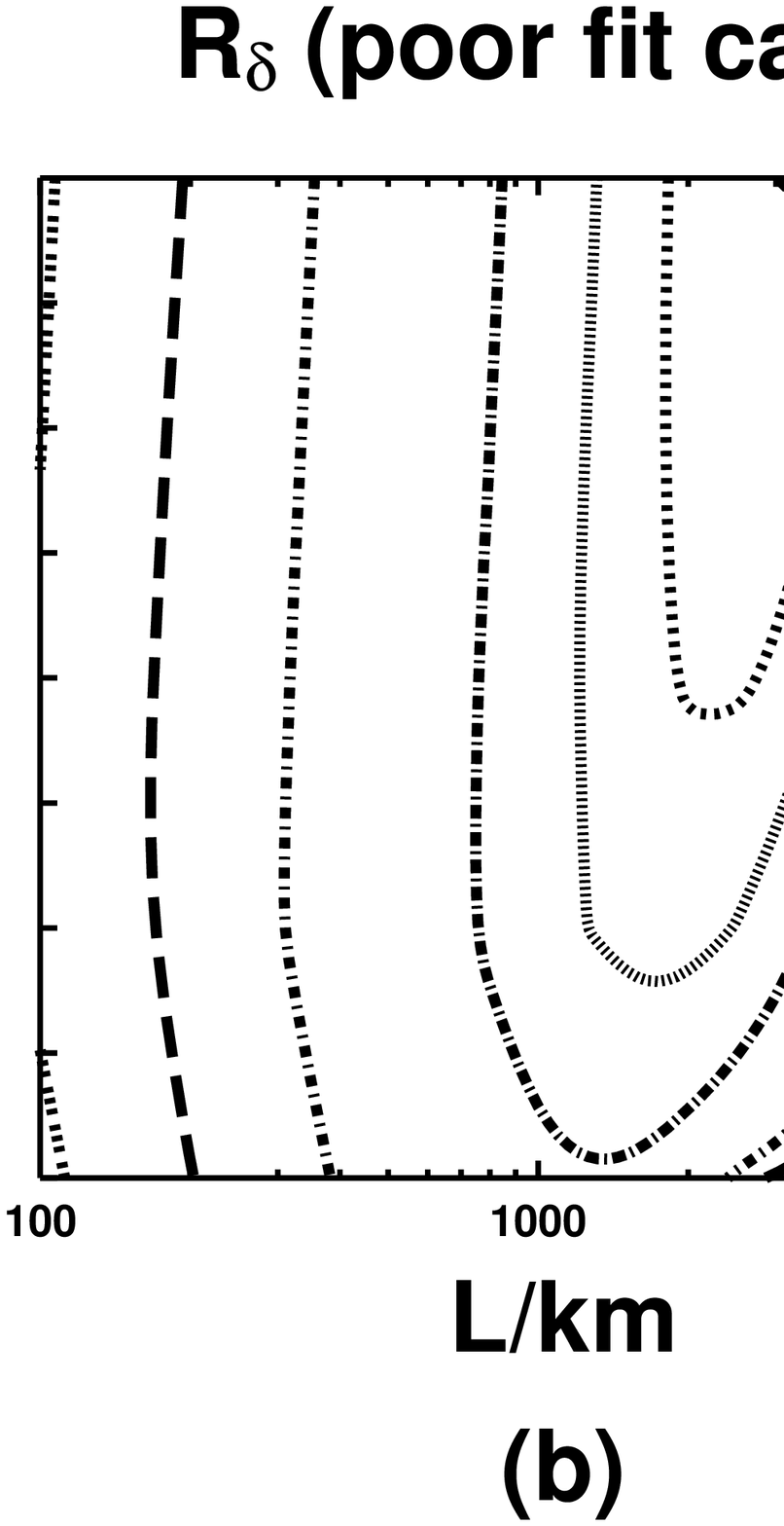,width=9cm}
\vglue -1.6cm \hglue 4.cm
\epsfig{file=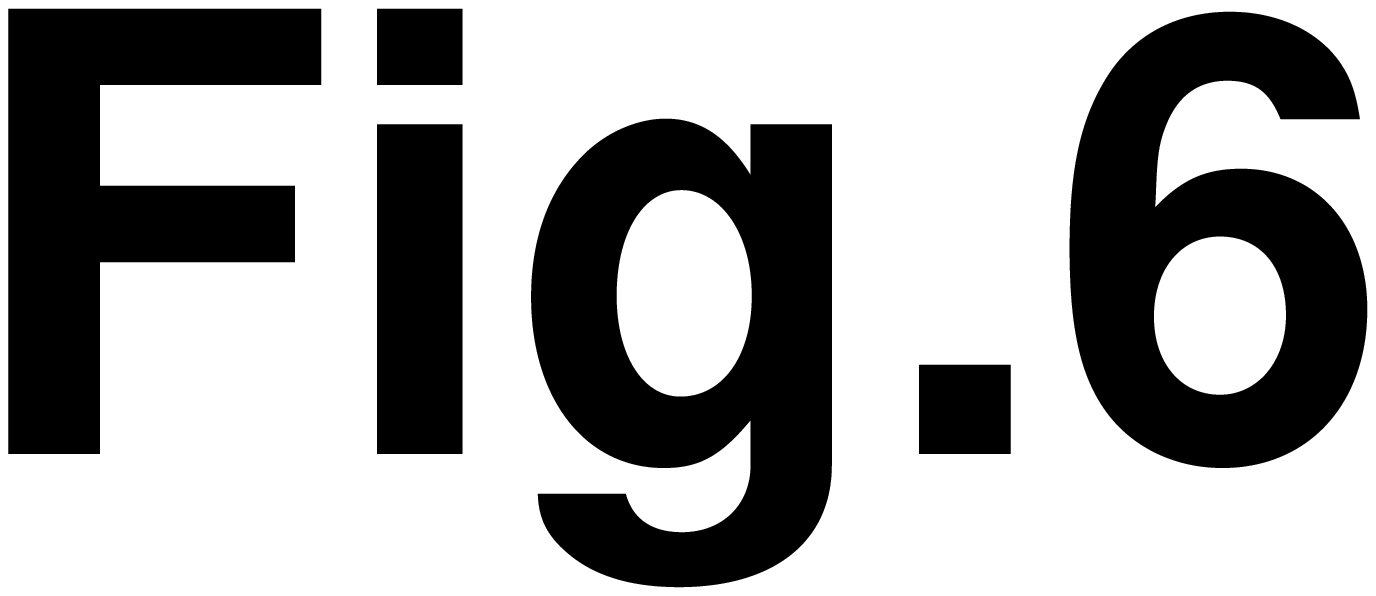,width=2cm}

If we take other set of parameters, e.g., 
$(\Delta m^2_{21},\sin^22\theta_{12})=(1\times 10^{-4}
{\rm eV}^2, 1.0)$, $(\Delta m_{32}^2,~\sin^22\theta_{23})
= (6\times 10^{-3}{\rm eV}^2,0.8)$ (Fig. 6b), which are still in the
allowed region of 99 \%CL of data of solar and atmospheric neutrinos, 
we have
\begin{eqnarray}
\max_{L,E_\mu} R_\delta \simeq 200\nonumber
\end{eqnarray}
for $\sin^22\theta_{13}=0.09$, $L\simeq 3000$km, $\delta=\pi/2$,
$E_\mu\simeq$50 GeV with 10$^{21} \mu$/yr$\cdot$10kt$\cdot$1yr.
I have also calculated the ratio $R_\delta$ for smaller value
of $\theta_{13}$, and found that the signal is optimized
for almost the same set of the parameters ($E_\mu\simeq$50 GeV,
$L\sim 3500$km), although the finite value of $R_\delta$ is
obtained as a limit of 0/0 for very small value of $\theta_{13}$ and
one has to make sure that we have a certain amount of events
to be conclusive.
If (\ref{eqn:r}) happens to be the case, then after running
the experiment for several years it may be possible to demonstrate
$\delta\ne0$.

\section{Summary}
In this talk I have discussed some quantities
(the sign of $\Delta m_{32}^2$, the magnitude of
$\theta_{13}$, the magnitude of $\delta$) which can be
measured at neutrino factories, and made efforts
to optimize the signals.  Of course all the measurements
would be impossible if $\sin^22\theta_{13}\ll 10^{-3}$, but
otherwise we may be able to determine the sign of
$\Delta m_{32}^2$, the values of $\theta_{13}$,
$\theta_{23}$ and the value of $\delta$ with our reference
values 10$^{21} \mu$/yr$\cdot$10kt$\cdot$1yr.  To measure $\delta$
it is necessary to know with great precision quantities
such as the cross sections $\sigma_{\nu N}$, $\sigma_{{\bar\nu}N}$,
the mixing angles $\theta_{12}$, $\theta_{13}$ $\theta_{23}$,
the mass squared differences $\Delta m_{21}^2$, $\Delta m_{32}^2$
as well as density profile of the Earth.
It is important to estimate how much the uncertainty on
$\Delta m^2_{ij}$, $\theta_{ij}$ and the density profile of the Earth
affects the reach of the experiment, particularly in the measurement
of CP violation, and this is a subject for future study.

\section*{Acknowledgments}
I would like to thank Yoshitaka Kuno and Yoshiharu Mori for
suggesting this subject and for discussions.
This research was supported in
part by a Grant-in-Aid for Scientific Research of the Ministry of
Education, Science and Culture, \#12047222, \#10640280.

\end{document}